# Essay: A path for the construction of a Muon Collider

Diktys Stratakis

*Fermi National Accelerator Laboratory, Batavia IL 60510 USA*

## Abstract

Muons are elementary particles and provide cleaner collision events that can explore higher energies compared to composite particles like protons. Muons are also far heavier than their electron cousins, meaning that they emit less synchrotron radiation that effectively limits the energies of circular electron-positron colliders. These characteristics open up the possibility for a Muon Collider to surpass the direct energy reach of the Large Hadron Collider while achieving unprecedented precision measurements of Standard Model processes. In this Essay, after briefly summarizing the progress achieved so far, I identify important missing R&D steps and envision a compelling plan to bring a Muon Collider to reality in the next two decades. A Muon Collider could allow for the exploration of physics that is not available with current technologies. For example, it may provide a way to study the Higgs boson directly or probe new particles, including those related to dark matter or other phenomena beyond the Standard Model.

*Introduction.*—Today, particle accelerators are one of the primary sources of discoveries about the fundamental nature of the world. The existence of the Higgs boson [1] was confirmed in 2012 at the Large Hadron Collider (LHC) at CERN, marking a significant scientific milestone. Although the LHC has fundamentally reshaped the landscape of high-energy physics, several questions remain: What is the role played by the Higgs in the origin and evolution of the Universe? What is the nature of dark matter? Why is there more matter than antimatter in the Universe? These are some of the myriad unanswered questions related to dark matter, the dark sector, neutrinos, and new phenomena within the Standard Model. Future colliders [2], reaching up to multi-TeV scales, can shed light on such questions, and several ideas have been proposed to push our understanding of the universe beyond LHC [3].

Most of the work conducted in the past decade has largely coalesced around linear [4] or circular electron-positron colliders [5] and circular proton-proton colliders [6]. Proton-proton colliders, like the LHC, can reach very high energies due to the proton's relatively large mass. However, protons are composite particles made of three quarks, so their collisions exchange only a fraction of energy and are messy. On the other hand, electron-positron colliders are precision machines because both particles are elementary without composite structure. As a result, when they annihilate, they convert all their energy into just a few products. But there is a problem when accelerating electrons in a circular accelerator: Electrons emit light and lose energy as they circulate the ring. This puts a limit on the accessible collision energy. These facts support the development of a third type of collider with the unique and remarkable capability to bridge the gap between energy and precision by achieving both simultaneously. This innovative and exciting concept is the Muon Collider (MuC) [7].

There are several advantages of colliding muons. Similar to electrons, muons are elementary particles, which means a muon-antimuon collision converts almost all the particles' energy into something new. Unlike electrons, however, muons can be accelerated to TeV energy scales with rings because they are roughly 200 times heavier and, therefore, emit about 2 billion times less synchrotron radiation. The practical consequences are noteworthy: A MuC ring with a circumference of 10 km could have the same potential as a 100 km proton collider ring, if proven to be feasible [8]. Additionally, the luminosity normalized as the wall plug power required by the MuC increases almost linearly with the center of mass energy, making it one of the most efficient machines above 2 TeV [9]. This is an important consideration in light of the global push for a more energy-efficient and sustainable future.

The concept of a Muon Collider (MuC) emerged in the 1960s [9]. Various design proposals were developed during the years [11-13], and a complete scheme for a MuC was available already in 2007 [14]. The Muon Accelerator Program (MAP) was approved by the U.S. Department of Energy in 2011 to advance MuC technologies [15], concluding in 2016. In 2021, the International Muon Collider Collaboration (IMCC) was established [16] to design a 10 TeV MuC and develop a related R&D program aiming for realization by 2040. In 2023, following the excitement generated in MuC during the Snowmass 2021 process [17], the Particle Physics Project Prioritization Panel (P5) recommended the development of a collider capable of achieving parton collision energies of 10 TeV [18]. Leveraging existing expertise and partnering with the IMCC could lead to construction readiness in the next two decades.

*Challenges and technology requirements.*—While a MuC has the potential to offer a precision probe of fundamental interactions in a smaller footprint as compared to electron or proton colliders, the short muon lifetime of 2.2 μs at rest and the difficulty of producing intense muon bunches with a high phase-space density generates some obstacles [19, 20]. Muon production schemes using protons appear more promising as compared to positron schemes [21, 22], due to their potential to generate large numbers of muons through tertiary production from protons hitting the target. The proton source must have a high intensity, and a very efficient capture of muons is required. Moreover, the newly born muons have large transverse and longitudinal emittances. The beams must be cooled by approximately a factor of 1000 in each of the two transverse phase spaces and by about a factor of 10 in the longitudinal using ionization cooling, the only known method fast enough. Acceleration to TeV energies requires the development of magnets that can ramp up the magnetic field quickly so that the muons can reach design energy before they decay. Finally, collider ring magnets and detectors must be shielded from decay products such as electrons and positrons. We conclude that mitigating the risk profile of a MuC requires the development of several key technologies and innovative concepts.

A conceptual scheme [23] for a 10 TeV center of mass energy MuC complex is illustrated in Fig.1. The proton driver (PD), the first piece of the machine, consists of a high-power acceleration section that produces a 5-15 GeV beam pulse, further intensified and shortened by an accumulator and a compressor ring, respectively. To achieve a desired luminosity of the order of $10^{35}$ cm$^{-2}$s$^{-1}$, a proton beam in the 2-4 MW range with a low repletion rate of 5-10 Hz is needed to provide enough muons downstream to satisfy this criterion. Short (~ 2 ns) bunches containing a few hundred trillion protons are directed onto a target, where pions are produced and then decay into positive and negative muons accompanied by neutrinos and antineutrinos. The *target* is a crucial component of the machine, as it must not only produce the highest possible number of secondaries but also be robust enough to withstand the shock thermal waves from intense beam pulses.

The initial muon beam occupies a relatively large phase space volume, which must be compressed by six orders of magnitude to obtain high-luminosity collisions. This phase-space reduction must be done within a time that is not long compared to the muon lifetime. *Ionization cooling* [24], illustrated in Figure 2, is currently the only feasible option for cooling a muon beam. It is achieved by reducing the beam momentum through ionization energy loss in absorbers while compensating for the momentum loss in the longitudinal direction through radio-frequency (rf) cavities. To be effective, absorbers should have a high product of radiation length and energy loss rate, with liquid hydrogen and lithium hydride being the best options.

Efficient cooling requires a small beam size at the absorber. Moreover, the minimum emittance that can be achieved is related to the beam size: A smaller size leads to a lower emittance. Therefore, the cooling channel contains solenoidal magnets that increase focusing strength along its length. A typical channel is about one kilometer long, containing over 1000 cells wherein a set of thousands of absorbers, solenoids, and accelerating rf cavities are closely interleaved. The focusing magnetic field starts at a relatively modest value of 2 T, and increases to several tenths of Tesla by the end of the channel.

Acceleration toward TeV scale energies will be achieved using a sequence of *Rapid Cycling Synchrotrons (RCS)*, typically three or four. In this scheme, the beam must pass through the same magnet at different stages and energies. The proposed configuration is a hybrid system that combines *fast-ramping magnets* with *static superconducting (SC) magnets*. During the injection phase, the fast-ramping magnets operate at full field strength but bend the beam outwards, partially compensating for the bending provided by the SC magnets. As the beam's energy increases, the fast-ramping magnets are adjusted downwards and then back up to align with the SC magnets until the beam can be extracted to the collider. It is crucial for the magnets to increase their field strength rapidly so that muons reach the desired energy before they decay. Ramp rates of ~300 T/s and a few kT/s are being considered for the high-energy and low-energy accelerating sections, respectively. The technologies currently under consideration for these magnets include normal-conducting (NC) magnets and High-Temperature Superconductor (HTS) magnets. In both scenarios, the maximum strength achieved by the ramping magnets is limited to less than 2 T. Consequently, the accelerator ring will be larger than the collider ring (see Fig. 1).

The collider ring should have the lowest possible circumference because the beam intensity decreases over time due to the random decay of muons. To address this, *high-field bending magnets* are preferred. For a 3 TeV collider, the design utilizes 10-11 T bending magnets made from niobium-tin (Nb3Sn), similar to those planned for the High Luminosity Large Hadron Collider (HL-LHC). Higher collision energies make it more challenging to bend the beam, necessitating stronger magnets. For a 10 TeV collider, researchers are considering 16 T dipoles similar to the ones for the proposed Future Circular Collider for hadron-hadron collisions at CERN. The electrons and positrons produced from these decays can hit the magnet, potentially causing radiation damage and heat load if not managed properly. Solutions being explored include the use of large-aperture bending magnets (~160 mm bore) combined with shielding materials, typically tungsten, placed within them. Encouragingly, the magnet shielding required for a 10 TeV collider is comparable to that for a 3 TeV collider, as the power per unit length of the particle loss remains similar [25]. Additionally, muons circulating in the MuC collider decay and generate neutrinos with a small solid angle that may lead to a large flux of neutrinos in the plane of the collider ring. These high-energy neutrinos have a non-negligible probability of interacting with material near the Earth's surface, generating secondary particle showers. One of the key challenges for the MuC is to minimize the radiation produced by these showers, ensuring it remains negligible.

*Timeline and R&D needs.*—The construction of the machine is expected to take around ten years, considering two scenarios: the *energy staging approach* starting at 3 TeV and progressing to 10 TeV as the technology matures, or the *luminosity staging approach* that begins at full energy with less efficient magnets, which will be upgraded over time.

Various options are being considered for the location of such a facility, including using the LHC tunnel [26] for a high-energy collider or constructing a new, smaller tunnel for the actual collider ring. Fermilab has also been considered as a potential site [8]. A detailed feasibility study is needed to assess these scenarios and identify the required R&D for adapting existing infrastructure for a MuC facility. To maximize the luminosity of a MuC, it is essential to integrate technologies related to magnets, accelerating cavities, the production target, and the proton source carefully. Each of these components must be optimized to reach peak performance. In this section, I will discuss the R&D needed in these areas to accomplish this goal.

*Design and simulations.*—To evaluate the performance of a 10 TeV MuC, it is essential to design and simulate its most challenging subsystems. One critical component is the *ionization cooling*. The conventional approach involves a 1-km-long cooling section that reduces both transverse and longitudinal emittances of the beam (6D cooling) [24], followed by a 200 m segment for transverse cooling only (4D cooling) [27]. The system was designed over a decade ago, and due to technological limitations, it assumed a maximum solenoid field of 30 T, resulting in a transverse emittance that was twice the desired value for achieving the luminosity goal. Nowadays, recent advancements in superconducting (SC) solenoids allow for fields up to 32 T [28], with the potential for over 40 T in the future at different high magnetic field user facilities [29]. Utilizing these higher fields could enhance focusing and reduce beam emittance. Future efforts should focus on integrating these technological advancements into the design. Additionally, conducting more extensive optimization studies could yield significant improvements. Recent advancements in machine learning (ML) optimization methods should also be leveraged to enhance the design and performance of the cooling channel.

The next critical subsystem of the machine requiring further study is the *acceleration.* As discussed earlier, a concept for achieving TeV energies with multiple RCS rings exists; however, we need to develop a self-consistent design of a complete RCS lattice. The goal, therefore, is to gather enough information in the design to confidently specify the maximum energy achievable with certain technology choices for the accelerating cavities and ramping magnets. Understanding the interaction between the beam and cavities is crucial in this respect. Due to its extremely large intensity ($2.7 \times 10^{12}$ muons), the beam can induce an electric field that partially counteracts the accelerating field, thereby reducing the net accelerating voltage it encounters. To assess and mitigate these collective effects, detailed tracking simulations are essential. This is a common focus within accelerator science, and there are expertise and tools of the community worldwide that should be leveraged.

The final subsystem is the *collider ring,* which must be designed to accommodate the desired collision energy of 10 TeV. This design will need to take into account the limitations imposed by the magnet technology (details provided in the magnet section below) as well as the control of neutrino radiation. To mitigate the effects of neutrino radiation, one approach is to install the ring approximately 200 meters underground so that the divergence of the neutrino flux leads to dilution of the resultant shower. While this strategy is adequate for a 3 TeV collider, additional measures are necessary for 10 TeV. One promising idea currently being explored by the IMCC is to place the beamline components in the arcs on movers. By periodically deforming the ring in small increments, the direction of the muon beam can be altered over time. Research has demonstrated [30] that a vertical deformation of the beam between -1 mrad to +1 mrad by the movers would result in an effective dose of 10 μSv/year, similar to the impact from the LHC. Moving forward, it is crucial to conduct numerical simulations to assess how these lattice deformations will impact the beam dynamics of the circulating muons. Evaluation of the neutrino flux mitigation system should also take into account engineering challenges and modify the design if necessary. This analysis is of high priority because the feasibility of the proposed neutrino radiation control strategy and the proposed locations for the collider depend on its outcomes.

*Proton Sources.*—Sources with the potential to become multi-MW proton drivers as needed for a MuC are available worldwide, such as the Spallation Neutron Source (SNS), the Proton Improvement Plan II linacs, and the European Spallation Source. However, further R&D is needed to identify modifications and upgrades to these facilities so that their current energy range (1-2 GeV) can increase to the desired (5-15) GeV range. For example, at Fermilab, this could be achieved [31] by extending the existing PIP-II linac or combining it with an RCS using current technology. An upgrade for the proton driver (PD) could be beneficial for accelerating both muons and neutrinos, to study charged flavor violation, and for dark sector experiments, fostering collaborations [32] within a broader scientific community.

The big challenges arise after accelerating the protons: Uniquely for a MuC, it is desirable to compress the protons into a very short bunch length of about 2 nanoseconds to maximize muon production. This will require the addition of a compression ring with finely tuned rf cavities to shrink the beam in time at the expense of increasing its momentum spread. This compression method is widely used in various accelerator facilities worldwide, but it does not achieve the short bunch lengths needed for the MuC. Before finalizing the design, the two key questions to address are: How much can we compress the beam before space-charge begins to hinder the process, and what are the strategies to mitigate this effect? Therefore, it is vital to focus R&D efforts on bunch compression in extreme space-charge. Facilities that can serve as close analogs to the MuC PD exist and can offer a valuable test bed for demonstrating these concepts. For instance, bunch compression [33] with MW-scale proton beams can be benchmarked at the SNS, while technologies for space-charge control and mitigation can be explored [34] at the Fermilab FAST facility.

*Targetry.*—The muon production and capture process for a MuC utilizes in-flight pion decay similar to neutrino experiments. However, the extremely short proton bunch for the MuC results in a density near $10^{15}$ cm$^{-3}$, which exceeds current target facility limits and risks mechanical and radiation damage to the target, degrading its quality. An R&D initiative must then be launched to identify materials that can withstand these conditions. High-performance computing combined with ML-optimizers can help determine the best target concept design, as well as the optimal compound mix to maximize muon yield while minimizing deterioration. Subsequently, it is essential to benchmark these findings in a dedicated beam irradiation facility for target research and development, such as HiRadMat at CERN. This would provide valuable feedback for the final design selection and prototyping. Preliminary studies [35] suggest that graphite, similar to the targets used at existing or planned neutrino sources, could be a viable option due to its exceptional resistance to thermal shock for a beam power of up to 2 MW. This is a welcome scenario for a MuC since it could open a path for synergies with ongoing targetry R&D efforts for neutrino experiments such as LBNF.

*Accelerating Structures.—* Accelerating structures, such as RF cavities, are crucial in the muon cooling channel, which contains thousands of normal-conducting (NC) cavities operating at frequencies between 200 and 800 MHz. To minimize losses due to muon decay, the cooling channel must be as compact as possible. As a result, the cavities must be positioned close to the focusing magnets, exposing them to strong magnetic fields from the adjacent solenoids. This field starts at a few Tesla (T) and reaches almost 10 T by the end of the cooling channel. Previous experiments [36] have demonstrated that this configuration complicates the operation of RF cavities, leading to significant surface damage and a reduced maximum gradient. Numerical calculations [37] suggested that this issue arises from electrons emitted from the wall and guided by the magnetic field to the opposing side. In this model, the surface damage and eventual breakdown are attributed to fatigue caused by cyclical strains from local heating due to these electrons. This suggests that using harder, lower-density cavity-wall materials would minimize this damage by allowing electrons to penetrate deeper, therefore reducing the local surface heating.. Experimental results [38] with a Cu-made cavity confirmed a three-fold reduction in the maximum achievable gradient when the magnet field was increased from 0 to 3 T, consistent with the model. When the same experiment was repeated with a Be-made cavity, which is harder and with a lower density, there was no gradient degradation, confirming that rf breakdown in the presence of magnetic fields is correlated to the thermo-mechanical properties of the materials used. This underscores the need for R&D aimed at developing materials that can better withstand these conditions. More resilient materials, such as aluminum (Al) or copper alloys like CuAg, have been proposed [39] and should be further explored. A dedicated test stand for these materials is also essential and should include one or more cavities of different materials along with a solenoidal magnet. This magnet must have a sufficiently large bore (ideally 350 mm) to accommodate the cavity while generating a field that mimics the conditions of an actual cooling cell, ranging from 7 to 10 T. A cryogenic Cu-made cavity that uses the technology proposed for a future linear electron-positron collider [40] could enhance surface hardness and thermal conductivity, reducing the likelihood of RF breakdown, and should be tested. The breakdown in the presence of magnetic fields is a challenge common to other areas, such as photoinjector design, advanced high-gradient accelerators, and klystrons, offering potential collaboration across fields.

RF cavities are also used in the acceleration section of the MuC. In this case, the RF cavities do not need to be positioned near magnetic fields, therefore, they can be SC. Compared to NC, SC technology offers several advantages, including lower power losses, allowing for larger beam apertures, higher accelerating fields, and less disruption to the particle beam. Current designs incorporate hundreds of 1300 MHz cavities per RCS operating at 30 MV/m, similar to those proposed for the International Linear Collider [41]. If collective effects become an issue, one potential mitigation strategy is to reduce the frequency to 800 MHz, which results in a larger cavity, less vulnerable to the interference from the incoming beam. Similar cavities are also being planned for the proposed electron-positron FCC (FCC-ee), suggesting a collaborative opportunity for superconducting radiofrequency (SRF) research and development. A unique challenge for muons is their constant decay, which generates numerous seed particles that can lead to multipacting, potentially degrading cavity performance. It is essential to evaluate this effect and optimize cavity shapes accordingly, as well as explore various surface preparation techniques. After selecting a suitable SRF cavity design for TeV acceleration, it must be validated with a prototype. Leveraging global SRF facilities will be crucial for producing and testing the cavity. Ideally, an intense proton beam [42] should be used for an *in situ* evaluation of the acceleration and collective effects.

*Magnets.*—To maximize the capture of muons, the production target is placed within a 20 T solenoid. The current baseline solution combines a Low-Temperature SC (LTS) magnet, generating a strong field of 15 T with a 2400 mm bore, complemented by a resistive Normal Conducting (NC) insert, providing the remaining 5 T with a 150 mm bore. The large bore of the LTS magnet accommodates shielding against heat and radiation coming from the interaction of the proton beam and the target. Recent advances in HTS magnets for fusion applications suggest that the LTS magnet can be replaced by a 20 T HTS magnet operating at 20 K [43]. This new approach offers several advantages: First, there would be no need for the resistive insert, significantly reducing power consumption. Second, due to its larger operating temperature (20 K) as compared to the LTS magnet (4.2 K), the HTS magnet can accept a higher heat load [44]. Therefore, the thickness of the shielding can be reduced, leading to an SC coil half in diameter and mass compared to the baseline solution. While this looks promising, further R&D is required to understand the radiation damage mechanism in HTS and to explore suitable HTS conductors and winding technologies for these types of magnets. These challenges mirror those in developing magnets for fusion devices, such as the central solenoids for tokamaks [45], indicating potential for collaboration in these areas. To prevent beam losses in the TeV accelerator, the RCS requires ramped magnets that achieve a field sweep of 3.6 T within a few 100 μs, which is 100 times faster than existing systems. Prototype tests suggest that resistive solutions, like pulsed iron magnets, achieving a maximum field of 1.5 T [46], and conventional copper coils [47], show promise, although higher fields could increase power losses. HTS coils [48] have also demonstrated a ramp rate of 300 T/s, matching the specifications for the last RCS ring, but fall short of the required field swing.

The next step is to test a full-scale ramped magnet that resembles the magnet intended for RCS acceleration and can reach the required peak field and ramp at the desired rate. The goals are to confirm the expected performance and to analyze engineering and cost aspects for the system. Additionally, extensive research is needed to design and develop a magnet power supply capable of handling tens of gigawatts, along with an effective energy recovery system, as no detailed designs currently exist. The magnets for the collider ring must meet two key requirements: they need to be strong enough to minimize the ring's circumference while also having a large bore to accommodate adequate shielding. Current studies on 10 TeV collider optics [49] indicate that the main arc magnets generate a steady state magnetic field up to 16 T in a 150 mm aperture, while the IR quadrupole magnets have a peak field of 20 T in a 200 mm aperture. Analytical studies [50] on the arc magnets show that Nb-Ti at 1.9 K cannot attain the desired field strengths. Nb3Sn at 4.5 K also falls short of performance requirements, providing feasible solutions only up to 14 T. REBCO, an HTS magnet, can achieve 16 T but at a 100 mm aperture. Using REBCO at temperatures between 10 K and 20 K is viable, as there are no concerns regarding the operating margin. Operating at a temperature higher than liquid helium is compatible with a large energy deposition in the coils, which leads to reduced shielding requirements and a smaller aperture, compatible with the estimate for REBCO.

While these analytical evaluations provide basic technology guidelines, further R&D is needed to advance to actual magnet design, engineering, and cost analysis. Research in this area aligns with the development of future hadron colliders, such as the FCC-hh, and can lead to transformative technologies with potential applications in areas like fusion.

*The muon cooling system.*—Among all components of a MuC, ionization cooling presents the greatest luminosity risk factor, estimated to be 3 or 4 orders of magnitude higher. Ionization cooling is a key component of the MuC: It enhances the beam brightness by repeatedly slowing the muons in absorbers and then re-accelerating them in RF cavities, all while maintaining the muons within a strong solenoidal field to ensure proper focusing. Efforts at the Rutherford Appleton Laboratory [51] and Fermilab [52] have demonstrated the underlying physics of ionization cooling, but the integration and technological challenges associated with this process remain the bottleneck for the MuC development. For instance, the operation of accelerating cavities close to solenoidal magnets may compromise the cryogenic performance of the magnet. Additionally, installing cooling absorbers in such compact assemblies can be problematic. Mitigation strategies to manage the forces exerted within and between the magnet coils need to be developed. Furthermore, RF power must be delivered to the RF cavities, which need to function efficiently within multi-Tesla magnetic fields. Finally, the instrumentation requirements and its operation under the cooling environment need to be assessed.

To understand and circumvent these risks, a Demonstrator facility [53], featuring ionization cooling cells that simulate a realistic cooling channel, is essential. This facility would aid in designing components and testing their integrated performance, paving the way for advancing cooling technology. A focused effort should be made in the coming years to conceptualize this demonstrator for testing cooling technology and identify potential hosting sites.

*Summary.—*A muon collider -MuC- would enable scientists to explore a wide range of high-energy physics phenomena. This includes the potential to uncover new physics beyond the Standard Model, investigate the nature of the Higgs boson in greater detail, search for dark matter candidates, and probe the fundamental interactions between particles with high precision. A MuC could be more cost-effective and energy-efficient than traditional proton colliders, thanks to the unique properties of muons. The development of a high-energy MuC has elements that overlap with other efforts in the field. For instance, the need for high-field magnets aligns with the ongoing research for future proton-proton colliders.

High-temperature superconducting magnets relevant for a MuC are also considered for fusion reactors and axion dark matter searches. Moreover, the development of MW scale targets will support other scientific advancements, as they are also necessary to provide the required beam flux for both long- and short-baseline neutrino programs. Finally, SC RF cavities, essential for muon acceleration, are in synergy with ongoing R&D efforts at CERN for the FCC-ee.

How can we embrace the *muon shot*? We must foster a collaborative, international effort aimed at addressing the existing research and development challenges, along with any new ones that may emerge. It will be essential to build a collaboration with a diverse set of expertise and specialization, experimentalists and theorists working together, and bridge the boundaries between fields.

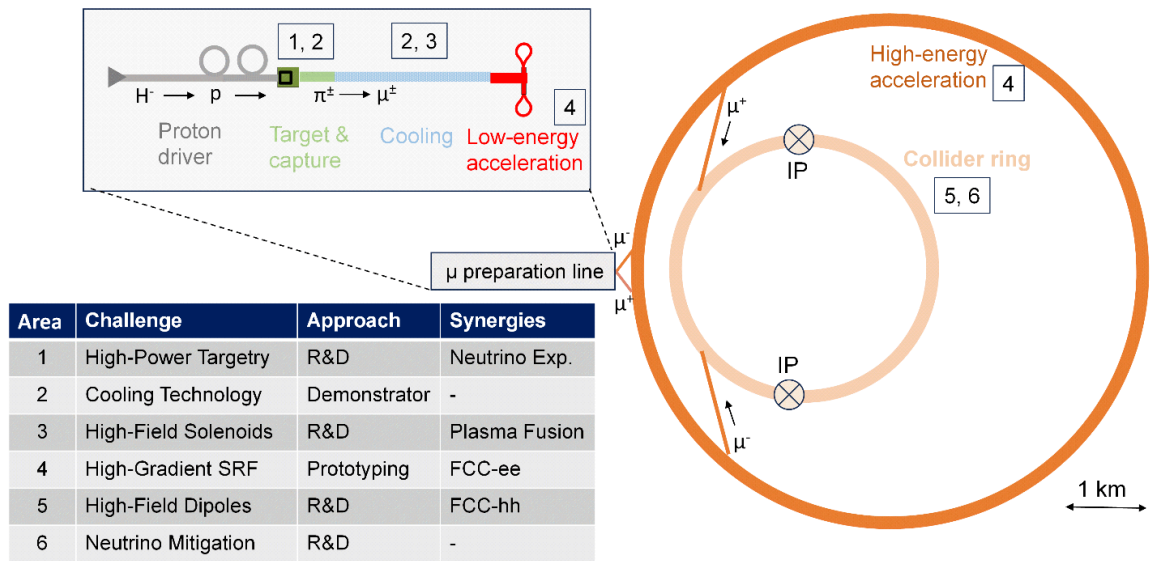

**Figure 1:** Illustration of a 10 TeV Muon Collider facility that consists of the following systems: A proton driver producing a high-power multi-GeV, multi-MW beam followed by an accumulator and a compressor ring that forms intense and short proton bunches. A target is inserted in a high-field solenoid to capture the pions and guide them into a decay channel where muons are collected. An ionization cooling channel that provides an emittance reduction for muons of both signs by nearly six orders of magnitude, and a muon accelerator stage that would deliver beams with energies up to ~150 GeV. A sequence of rings accelerates the beam to the multi-TeV range before injecting it into the collider ring. The ring is anticipated to support two detector interaction regions. The table on the left shows the main challenges, the immediate next steps, and possible synergies with other efforts.

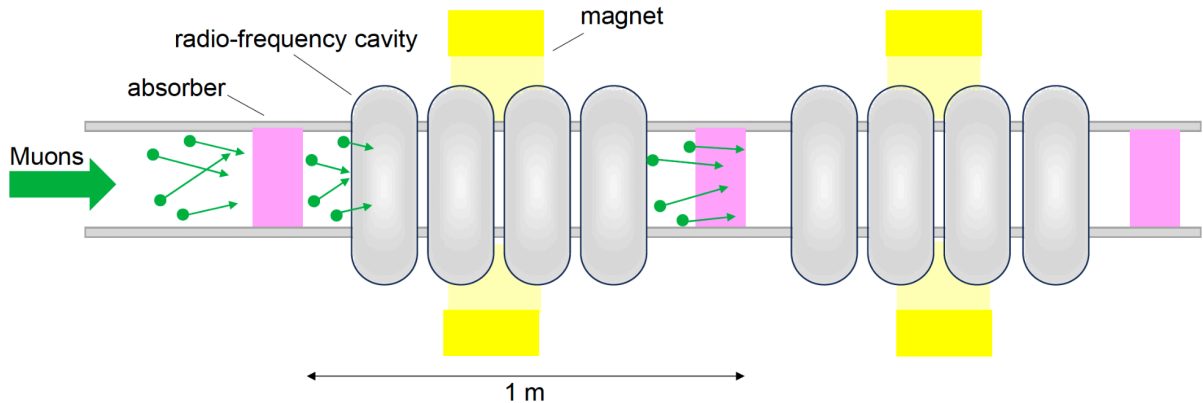

**Figure 2:** A schematic illustration of the ionization cooling channel needed for a Muon Collider. Ionization energy loss in all three planes occurs in absorbers, whereas the radio frequency cavities

restore only the longitudinal component. Repeating the process many times reduces the transverse emittance. The channel is immersed in superconducting solenoids to maintain the beam focusing.

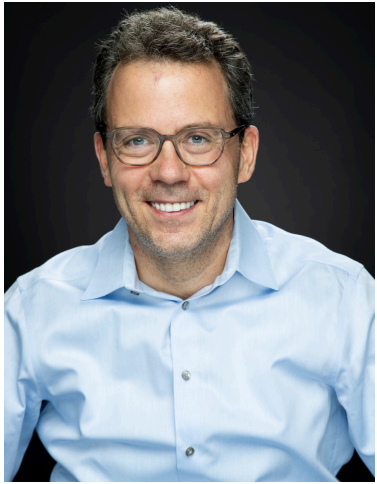

Bio: Diktys Stratakis is a scientist and group leader at the Fermi National Accelerator Laboratory in the USA. He holds a Bachelor of Science in Physics from the University of Crete in Greece, a Master of Science in Physics from the University of Florida, and a Ph.D. in Electrical Engineering from the University of Maryland. His research focuses on high-energy physics, with a particular emphasis on both experimental and theoretical accelerator physics. This includes exploring advanced accelerator concepts to enhance the performance of particle accelerators and developing diagnostic tools for their operation. For about 15 years, Stratakis has been dedicated to advancing the concept of Muon Colliders. He has made significant contributions to the development of muon cooling systems and the experimental demonstration of longitudinal ionization cooling. Stratakis has delivered over 100 invited talks, published more than 60 peer-reviewed articles, and is an elected IEEE Senior Member.